**Is India's Unique Identification Number a legally valid identification?**

Anupam Saraph[1],  Lalit Kathpalia[2], Anab Kidwai[3], Aniruddha Joshi[4]

## Abstract

A legally valid identification document allows impartial arbitration of the identification of individuals. It protects individuals from a violation of their dignity, justice, liberty and equality. It protects the nation from a destruction of its republic, democratic, sovereign status. In order to test the ability of an identification document to establish impartial identification of individuals, it must be evaluated for its ability to establish identity, undertake identification and build confidence to impartial, reliable and valid identification. The processes of issuing, using and validating identification documents alter the ability of the document to establish identity, undertake identification and build confidence to impartial and valid identification. These processes alter the ability of the document to serve as proof of identity, proof of address, proof of being a resident, or even the proof of existence of a person. We examine the ability of the UID number to serve as an identification document with the ability to impartially arbitrate the identification of individuals and serve as proof of identity, address, and demonstrate existence of a person. We evaluate the implications of the continued use UID system on our ability to undertake legally valid identification ensure integrity of the identity and address databases across the world.

**Keywords:** UID, Identity, Identification, Authentication, Corruption

## The Unique Identification Number

The Unique Identification Authority of India (UIDAI) was established in January 2009 through an executive order of the Government of India, as an attached office to the Planning Commission of India. It was established to issue a unique identification number

[1] Anupam Saraph (anupam.saraph@sicsr.ac.in) is Adjunct Professor, Symbiosis Institute for Computer Studies and Research and former Advisor to Government of Goa

[2] Lalit S Kathpalia (director@sicsr.ac.in)  is Director, Symbiosis Institute of Computer Studies and Research

[3] Anab Kidwai  (anab_kidwai@iiml.ac.in) is Research Associate, Indian Institute of Management Lucknow, and PhD scholar at Symbiosis Centre for Research and Innovation (SCRI)

[4] Aniruddha Joshi (joshiag@pun.unipune.ac.in) is Director, Centre for Information and Network Security, Savitribai Phule Pune University

(UID or Aadhaar) to all Indian residents that can eliminate duplicate and fake identities, and be verified and authenticated in an easy, cost-effective way (UIDAI 2010: 1).

The UIDAI describes itself as the regulatory authority to manage a Central Identities Data Repository (CIDR), issue UID numbers, update resident information, outsource access to the CIDR to agencies to enable them to authenticate the identity of residents as requested, and share KYC information about a resident with any licensed requesting entity, on approval of the concerned resident (UIDAI 2010: 2).

In order to enrol residents into their database, the UIDAI worked with partner agencies such as central and state departments and private sector agencies who were appointed as 'Registrars' for the UIDAI. As shown in Figure 1, the registrars processed applications from residents for being allotted a UID number, submitted data to the CIDR in order to de-duplicate resident information and received the UID number allotted to the submitted data. These Registrars could either enrol residents directly, or could appoint private agents or agencies as enrollers. The enrolment agencies interfaced with individuals who desired a UID numbers. The UIDAI permitted the registrar's flexibility in their processes, issuing cards, pricing, expanding KYR (Know Your Resident) verification, collecting demographic data on residents for their specific requirements, and in authentication. The UIDAI only required that demographic and biometric information submitted to it, for allocation of a 12-digit random number or the UID number, would be uniform (UIDAI 2010: 2-3).

The UIDAI has also licensed private agencies, who serve as service providers, who can query the UIDAI's CIDR to authenticate demographic or biometric information submitted by the partner agencies. Through these agencies, the UIDAI offers an online authentication, where user agencies can make requests to UIDAI to compare demographic and biometric information submitted by them with that associated with the UID number record stored in the UIDAI database (UIDAI 2010). On biometric verification of the biometric submitted by an agency with that associated with the UID, and considering that as consent to transmit eKYC information of the UID number, such an agency can obtain eKYC information associated with a UID number from the UIDAI (UIDAI 2012f).

*Authentication* of information submitted to the UIDAI to compare with that associated with an UID number in UIDAI's CIDR is being considered as *identification* of the citizen. Hence, the UID number is being used to replace existing know your customer/ resident (KYC/ KYR) requirements, to create citizen databases like the National Population Register (NPR) and voter lists, to extend rights of citizenry (like voting, obtaining passports) to residents, to facilitate access to records with the same primary key for every database in the country and even to open bank accounts and do financial transactions. The UID number has poised itself as the ubiquitous primary key of private and public databases.

The UID number, issued by the UIDAI, is being widely regarded in India as proof of identity, resident status, as well as address. Both public and private institutions are doing away with their existing practices of collecting KYC (Know Your Customer) information in favour of identifying citizens and clients through their UID numbers. With function creep linking the UID number to an increasing number of public and private databases, and replacing existing KYC practices, it is essential to exact the UIDAI's claim that the UID number will provide and prove unique identity.

Without defining the ability of the Aadhaar number to serve as an identification document, section 4(3) of the Aadhaar (Targeted Delivery of Financial and Other Subsidies, Benefits and Services) Act, 2016declares "An Aadhaar number, in physical or electronic form subject to authentication and other conditions, as may be specified by regulations, may be accepted as proof of identity of the Aadhaar number holder for any purpose". There is urgent need, therefore, of evaluating the UID number for its ability to establish identity, undertake identification and build confidence to impartial, reliable and valid identification There is also need to assess the consequences of using the UID widely, as is happening across India.

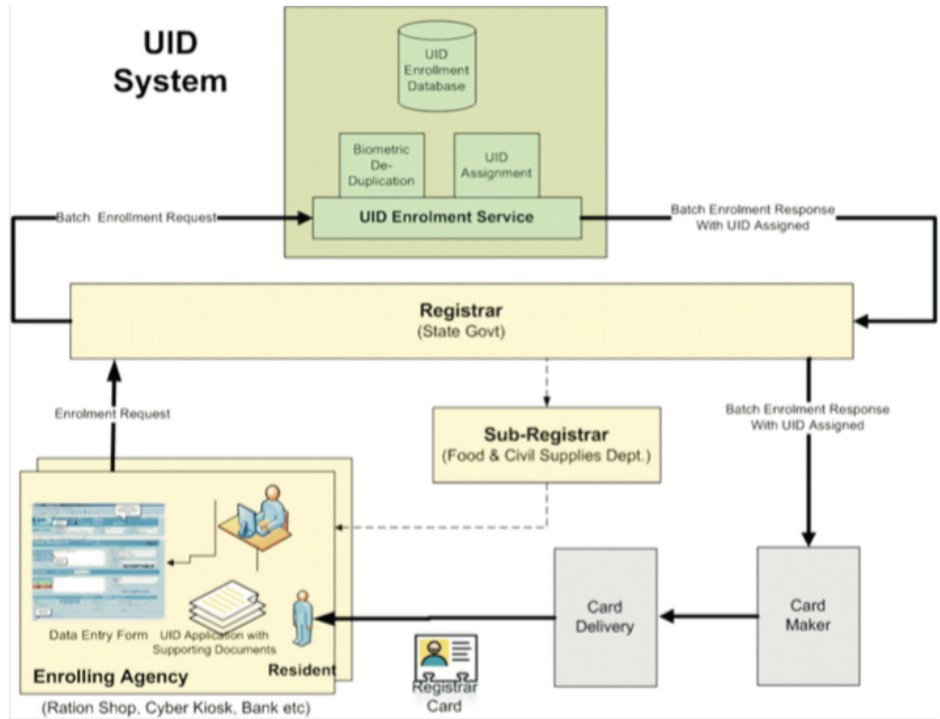

Figure 1: UIDAI process for Issue of UID Number (*Source: UIDAI 2010*)

**Methodology for evaluation of the UID to serve for identification of individuals**

Any legally valid identification document would need to allow impartial arbitration of the identification of individuals. In order to address the question if the UID allows impartial arbitration of the identification of individuals, we examine the ability of the UID to establish identity, undertake identification, and build confidence to reliable, impartial and valid identification.

In order to establish identity with an identification document, the process of creation of that identification document would require the identification of the individual, who is to be issued an identification document, and require a certification of the identification document by the issuer of the document (Saraph 2014). Without any identification of the individual, and any other required attributes like resident status, address or existence, an identity document cannot serve as a reliable, impartial and reproducible basis for establishing identity. Without the issuer of the document taking responsibility of certifying the identity and other contents of the document, an identity document cannot serve as a responsible and legally enforceable basis for establishing identity, particularly for use by third parties.

In order to establish identification of individuals using an identification document, the identification document needs to be authenticated and the process of identification defined to establish the identity and responsibility of identification. Furthermore to protect misuse of the identity documents for identification requires that the issuer provide a means to those identified to regulate and restrict the use of the identity documents (Saraph 2014). Without authentication of the identification document used to establish identity, genuineness of the identification cannot be established. Without defining the identification process and the parties responsible for identification, there is no identification. In the absence of a means to regulate and restrict the use of the identification document, misuse or misidentification cannot be prevented.

Finally in order to build confidence in the identification process, requires an independent third-party audit of the identity or address documents issued and the logs of the use of these documents. It also requires any identity or address attributes that change be added to the identity document, not overwritten (Saraph 2014). Without a verification and audit of the processes that generated the identification document, and the database of such documents itself, little trust can be placed in such identification documents. Without a means to see every change made to the identification document they do not inspire confidence in the validity of the identification that they can enable.

We use these criteria for the creation, use, and validation of identification documents to evaluate the process of creation of the UID, use of the UID for identification, and the validation of the UID.

We evaluate the implications of the continued use of the UID system on our ability to undertake legally valid identification and ensure integrity of the identity and address databases across the world.

Figure 2: UIDAI Enrollment Process as described by UIDAI
*(Source: UIDAI 2012a)*

**Creating a UID Number**

*Does the UIDAI ensure the identification of the individual?*

The UIDAI appointed State Government/ Union territory, public sector undertakings and other agencies and organizations, who interact with residents in the normal course of implementation of some of their programs, activities or operations as its Registrars. They in turn, appointed their departments as Sub-Registrars. These sub registrars in turn, appointed Enrolling Agencies. The Enrolling Agencies in turn, collect and submit demographic and biometric data for assigning a UID number and inclusion in the records for the UIDAI database. The enrolment process is described by the UIDAI schematically

**ER 4.5    Capture Demographic and Biometric Data and Ready for Transfer Sub Process Flow**

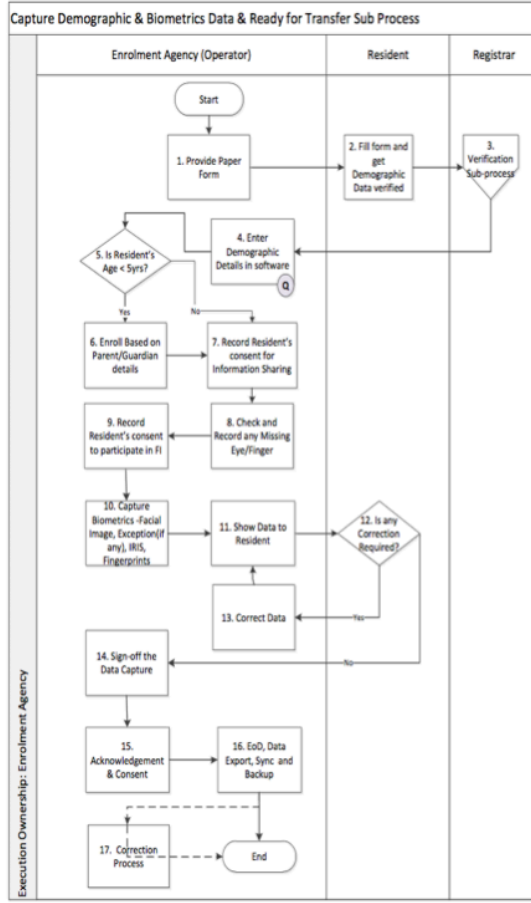

**ER 4.7    Verification Sub Process Flow**

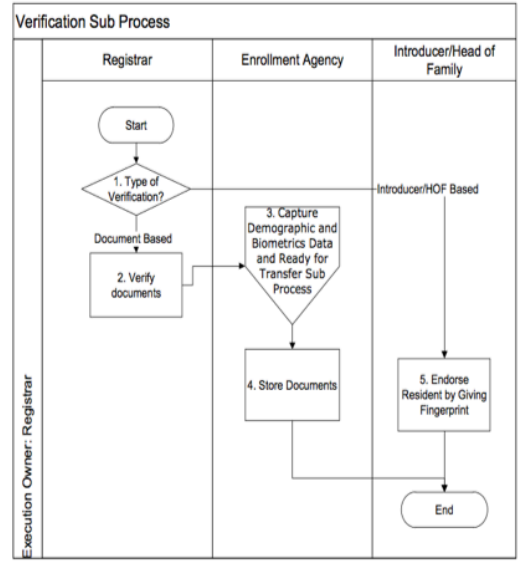

in Figure. 2 (UIDAI 2012a). While the creation of an identification document requires identification of the individual who is to be issued an identification document, there is no step in the UIDAI's process to establish the identity of the individuals, the resident status, or even verify their addresses.

The UIDAI enrolment process has a step to verify the documents submitted by a resident wishing to enrol for a UID number. The Registrar are given the task of verifying the authenticity of the documents, *not* the identity or address of the individual submitting the documents (UIDAI 2012 a). In practice Registrars delegated the verification of the documents to a supervisor appointed by the Enrolment agencies (UIDAI 2011, Union

Bank of India 2011, Allahabad Bank 2011, Bank of Maharashtra 2012, Oriental Bank of Commerce 2013).

Identification requires confirming that the information provided by the individual matches the individual being identified. This means the identifying party has to be co-present and legally responsible for the identification. This is essential to establish the identity match of the person present and the information provided. The UIDAI enrolment process does not require the identity match; at best, it merely requires the submitted documents to be verified (UIDAI 2012a).

From the UIDAI's documented enrolment process it is evident that the UIDAI only takes the step of verifying existing documents as opposed to identifying an individual in order to issue a UID number. The enrolment agencies have at best verified documents shown to them at time of enrolment and associated biometric information with the documents. This is not an identification of any individual. The UIDAI presumes that the associated biometric uniquely identifies the person associated with a document used for verification in future authentication using biometrics. Even if this were so, it has not established the identity of the biometric source associated with the documents.

This procedure does not eliminate the ability to use multiple biometric sources and associating them with demographic information from "verified" documents creating ghosts or duplicate UID numbers. Several organized criminals generating such identities have been busted across India (MidDay 2017).

As each Registrar had already issued the identity documents being verified, the Registrars could have identified the individuals based on documents already issued by them, in their possession, and certified identification. Alternately, each Registrar could have simply provided a database of those already issued their documents. The existence and identity of these persons could have been confirmed at the time of any transaction of each document holder with the Registrar. If required biometrics could have been captured after identification to build a database of those biometrics that occurred multiple times in the database in order to do a physical verification that these are distinct individuals.

Furthermore in a response to a query under the Right to Information Act, the UIDAI indicated that the information about the documents used to issue a UID number is not compiled/not available (RTI 2018). There is no basis, therefore, to hold it as a valid or reliable identification document that establishes proof of identity, address, resident status or even existence of any individual assigned a UID number.

*Does UIDAI certify the identity and address associated with the UID number?*

According to the UIDAI, its purview is limited to the issue of unique identification numbers linked to a person's demographic and biometric information (UIDAI 2010). The UID number allotment letter is not signed by any government official. It does not certify that any identification process has happened or even the identity and address verification of anyone. The Ministry of Finance in a reply under the Right to Information has confirmed that it has no information on the designation of officials who certify identity, address or existence of persons who are assigned a UID number or the evidence of the auditability of such claims. The UIDAI in response to a similar question in 2014 listed out the enrolment process (RTI 2014) and indicated that no such role is assigned to UIDAI officials (RTI 2018). Furthermore, no official of the UIDAI or any government department affixes a signature and seal on the UID card certifying the identity or address associated with the number.

Certification of the identification process is essential to identify the person and agency responsible for the issue of the identification document and establish their legal liability for the identification with the person. The UIDAI's processes do not require or result in any person or institution certifying the identity or being liable for (mis)identification. Furthermore, publicly available documents on the UIDAI website do not indicate any individuals who are responsible for certifying the identity of any person assigned a UID number or that the identification happened. It also follows that there is no one who even certifies the address, resident status, or even that there is a real person associated with the number or the demographic or biometric data associated with the data. There is no step in the enrolment process (Figure 2) or allocation of a number (UIDAI 2012 a) where anyone is required to even certify that the enrolment process actually happened as described or that the demographic and biometric information collected was as per the UIDAI process and belongs to the same person who is seeking to be enrolled.

Figure 3: UIDAI's response to RTI
*(Source: RTI 2018)*

The UID number, therefore, does not provide a reliable, impartial and reproducible basis for establishing identity. It also cannot serve as a responsible and legally enforceable basis for establishing identity, particularly for use by third parties.

**Using a UID Number**

*Does the UIDAI provide a means to authenticate the UID number?*

The UID number is a unique 12-digit random number assigned to a packet of demographic and biometric information submitted by private enrolment agencies paid for

No. 4(4)/57/ 330/2017-E &U
Government of India
Ministry of Electronics and IT
Unique Identification Authority of India
*******

2nd Floor, Tower-I, JeevanBharti Building,
Connaught Circus, New Delhi-110001
Dated: 22 Jan, 2018

To,

Shri Anupam Saraph,

Sub:- Information under RTI act 2005 :reg

Sir,

Please refer your RTI application Rgn. No. 50022 dated 05/01/ 2018 and UIDAI letter dated 11th Jan,2018, the information as requested is furnished as under :-

**Point 1:** The names and designations of the UIDAI officials certifying the identity, address, date of birth, resident status or existence of any individual and signing the certified record.

**Reply :** No such role is assigned to UIDAI officials. You may elaborate your query.

**Point 2** Whether the CIDR when queried with only the biometric data of an individual of an individual pulls up a unique record.

**Reply :** No.

**Point 3:** The documents used for POI, POA, DOB and POR along with number of Aadhaar issued by each combination of these documents in each district.

**Reply :** The information is not compiled/ available.

Point 4: A complete list of operators and supervisors registered with the UIDAI from inception with their complete address, enrolment agency and registrar.

**Reply :** The information sought is personal information of third parties and can not be provided under section 8(i)(j) read with Section 11 Sub section 1 of RTI Act 2005.

If you are not satisfied with the reply, you may appeal to the Appellate Authority, UIDAI within 30 days from the receipt of this letter to Sh. Narendra Bhooshan, Deputy Director General, Unique Identification Authority of India, Jeevan Bharati Building, New Delhi-110001

Yours faithfully,

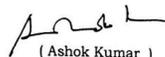

( Ashok Kumar )
Assistant Director General & CPIO,
Tel No. 23466840

Copy to :- Nodal Officer, UIDAI : w.r.t. case No. 2176 dated 8th Jan,2018.

---

each record. Like any other primary key in a database, the 12-digit number merely identifies each record uniquely (UIDAI 2012c).

The UIDAI has created a web service for checking if a random 12-digit number is actually an already assigned UID number or not.[5] Querying a 12-digit number on this website returns a message stating if the number is a valid UID number or not, and also returns the age band, gender, state of domicile, and the last 4 digits of the mobile number, if the number is valid. Otherwise, it offers an error message stating that the number is not found but does not conclusively deny the absence of the number. Thus, with access to the UIDAI portal, it can be possible to establish if the 12-digit number being queried exists in the UIDAI database or not.

---

[5] https://resident.uidai.net.in/aadhaarverification.

A photocopy of the UID card is however extensively used as an identification document. In the absence of querying the UIDAI portal, however, there is no way to understand by looking at the number or the UID card if it is an authentic UID number or card issued by the UIDAI. There is nothing in the UID numbering system that informs if a particular 12-digit number can be a genuine UID number or not. This excludes the digitally challenged from recognizing the validity of the number.

Furthermore while the UIDAI portal may confirm the UID number is issued by the UIDAI, it has no means to allow those without access to the CIDR to confirm that the demographic info associated with the photocopy is that which is associated with the same number.

Worse the UID contains at least three information channels. The physical photocopy, the QR code and the online data associated with the number. Any one can be altered and allow the authentication of the number as being valid. There is no means to authenticate the information associated with the number as being valid and the one captured by the UIDAI in its processes.

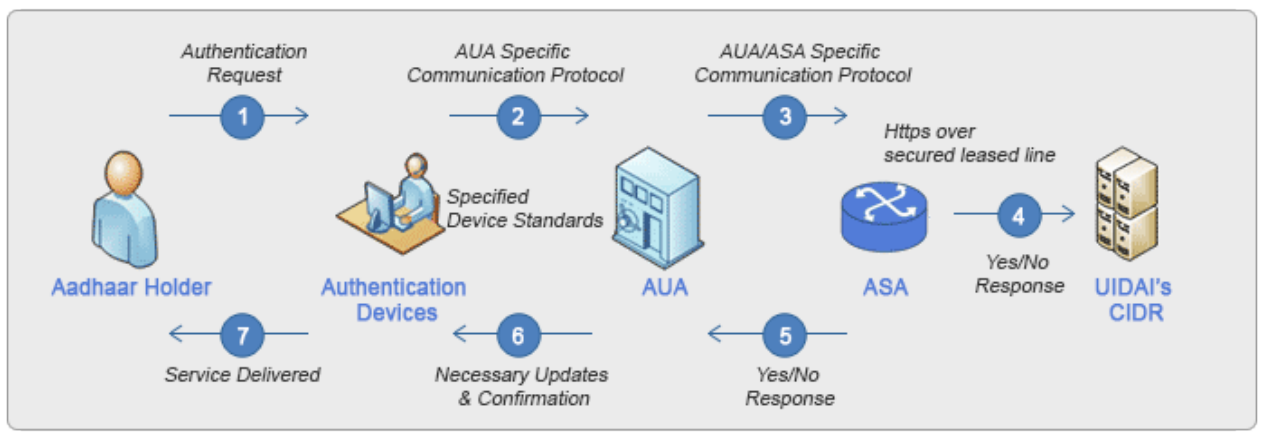

Figure 4: Aadhaar Authentication Ecosystem[6]

An Authentication User Agency (AUA)[7] can query the UIDAI database through an Authentication Service Agency (ASA)[8] to authenticate a demographic or biometric field

---

associated with the UID number (UIDAI 2012f). The AUA or ASA can only confirm or deny the value of the demographic or biometric field associated with the UID number in the UID database as being identical or different from the one queried. Unless both the demographic and biometric information is queried, a person can easily use their own number and biometric and provide someone else's demographic information.

UIDAI's authentication is one factor, static query to a remote database that cannot distinguish between a stored demographic or biometric query from a real person being present. Importantly, in UIDAI's authentication, there is no person who is responsible to establish identity. An identification process, in contrast, involves multiple, dynamic comparisons by a person who is responsible for establishing identity and is co-located and present with the person being identified. UIDAI's authentication is, therefore, distinct from identification where an individual must be identified based on information provided.

By failing to define identification in the Aadhaar (Targeted Delivery of Financial and Other Subsidies, Benefits and Services) Act, 2016, the UIDAI treats authentication as identification. Particularly section 4(3) of the Act incorrectly considers the UID number subject to authentication and other conditions, as proof of identity of the UID number holder.

For example, the UIDAI states that it offers an authentication service that makes it possible for residents to authenticate their identity biometrically (UIDAI 2012f). There is no service of identification of individuals but of authenticating data submitted by a user agency with the data stored in the UIDAI database. The UIDAI implies that individuals to have been identified when merely data submitted to the CIDR for comparison has been matched. At no time is the UIDAI in a position to verify from authentication if it is the same individual or even if the individual was present and it was not stored data, from a machine or database, that was submitted to authenticate a transaction.

The UIDAI's CIDR responds to five types of authentication requests:
- In Type 1 authentication the UIDAI authenticates an UID number with demographic attributes (name, gender, date of birth, etc).
- In Type 2 authentication the UIDAI authenticates access to a mobile number associated with the UIDAI number using a One-Time-Password (OTP) delivered to the mobile number.
- In Type 3 authentication the UIDAI matches either iris or fingerprint submitted with one associated with the UID number.
- In Type 4 authentication both the mobile and biometrics (either iris or fingerprint) are matched, and
- In Type 5 authentication mobile, fingerprint and iris together fare matched.

None of these acts of authentication, however, constitute *identification* of the individual. They merely authenticate (or match) the data submitted with that associated with the UID number. There is no way to use the number as a basis of identification as it does not provide a means to identify the individual present or even to require the presence of an individual who is authenticated.

Authentication provides access to anyone with a key. Identification requires a responsible party to ensure access to the identified individual is never denied.

If identification and authentication are treated synonymously, it is possible for an identity document to be used by third parties to impersonate others. If identification of individuals is skipped, impersonators can submit documents that may be authentic but belong to someone else. If authentication of documents is skipped impersonators can submit documents that are not issued by the certifying party yet claim to do so. By treating identification and authentication synonymously, the use of UID numbers allows transactions without undertaking identification and even permits impersonation.

*Does the UIDAI provide a scheme to regulate and restrict the use of the UID?*

Section 7 of the Aadhaar (Targeted Delivery of Financial and Other Subsidies, Benefits and Services) Act 2016[9] restricts the use of the UID number to authentication. In section 2(c) the Act defines authentication as a process by which the UID number along with demographic information or biometric information of an individual is submitted to the Central Identities Data Repository (CIDR) for its verification and such repository verifies the correctness, or the lack thereof, on the basis of information available with it. Section 29 places restrictions on sharing information to that specified in the Act (for authentication purposes only) in the manner specified by the regulations. These prohibit the retention, storage, display and publication of the UID number and associated biometric.

In practice, the UIDAI does not have any mechanisms to restrict the use of Aadhaar numbers to authentication, to discover any use of Aadhaar numbers for uses other than authentication. It does not have any mechanisms to even prevent or discover retention, storage, display publication of the UID number in violation of the Act. Such use and retention of the Aadhaar is widespread as has been reported where several hundred million of UID numbers were exposed in violation of the Aadhaar Act (Sinha 2017).

According to the UIDAI, Enrolling Agencies may store the information of residents they enroll (UIDAI 2010: 4). The registrars, sub-registrars or enrolment agencies will also be free to print/ store the biometric collected from the applicant on the issued card (UIDAI

2010: 15). Once information is retained by third parties for any use they may make of it, the UIDAI has lost any ability to restrict or regulate its distribution or use.

Section 3(2) of the Aadhaar Act (Government of India 2016) requires that the enrolling agency shall, at the time of enrolment, inform the individual undergoing enrolment, the manner in which the information shall be used and the recipients with whom the information is intended to be shared. Section 8(3) of the Aadhaar Act requires that a requesting entity shall inform the individual submitting his identity information for authentication, the nature of information that may be shared upon authentication and the uses to which the information received during authentication may be put by the requesting entity.

While the UIDAI allows those with a verified mobile number associated with the number to block or unblock access to the biometric associated with their UID number[10], this only restricts access through biometric and not any other authentication mechanisms like the mobile used to block and unblock the biometric.

The UIDAI does not provide any mechanism to ensure, that as per the Act, the individual at enrolment or authentication is aware of the nature of information shared, the use to which it may be put, the recipients of the information (UIDAI 2012a, 2012b, 2012d). The UIDAI provides no mechanism to the individual at enrolment or authentication stage to permit, prevent or restrict the use of the UID number to access any data associated with it. Despite requirement in the Act, the UIDAI does not provide any means to allow the individual to block or restrict access of data associated with the UID number to specified parties, for specific periods or specific purposes. It does not provide the individuals, any means to delete their UID number.

Despite the need to protect the individual from use of the UID for purposes beyond those consented, period beyond the consent or by persons beyond those who were granted access, the UIDAI does not make any logs, or provide any access to such logs, of the use of the UID number to seed, link, access or modify data associated with it available to anyone. It does not provide any means to know which, if any, third parties, particularly agencies who are required to use it to provide benefits, have used their UID number for any transactions linked with UID number to deliver benefits. It is impossible to know who accessed the data associated with the UID number. It is impossible to know where all the UID has been seeded or used to open bank accounts, obtain passports or drivers licenses etc.

---

10 https://resident.uidai.gov.in/biometric-lock

The UIDAI has destroyed the traditional responsibility of identification from persons and organizations providing services or entitlements. While doing so the UIDAI has not taken any responsibility for the business process that depends on the ability to respond humanely to identification challenges as well as identity frauds. This actually breaks most of the existing business processes, making them fail for the individual. It is therefore, important for the individual to actually know and be able to hold responsible the UIDAI for failures and misuse of the UID. This can only happen if there is a monthly statement, like a credit card statement every month on the use of their credit card, on the use of their UID to every UID owner.

Section 23(g) of the Aadhaar Act allows the UIDAI to omit or deactivate a UID number or data associated with it. So far, in exercise of these powers, 85,67,585 Aadhaar numbers have been deactivated by the UIDAI (Yadav, 2017). The UIDAI, however, has no publicly shares procedure to deactivate the UID number.

From the above, we can conclude that while the UIDAI can block a UID number by deactivating or omitting it, it has no mechanism to regulate, control, restrict, track or trace the nature of information shared, the use to which it may be put, or the recipients of the information.

Without authentication of the UID number and the information associated with it that is used for any process, the genuineness of any process based on it cannot be established. Since the UIDAI does not define or undertake any identification process, and does not designate any persons responsible for identification, there is no identification that is possible using the UID. In the absence of a means to regulate and restrict the use of the UID, misuse or misidentification, when using the UID, cannot be prevented.

**Validating a UID Number**

*Has there been an audit of the identity and demographic data associated with the UID number? Can there be one?*

In order to establish trust in the data associated with the UID number, it is necessary for auditability of the data associated with the number. Such auditability would require that the third party auditing the number is able to verify -

- that the data associated with the number actually corresponds to a real person,
- the same person had actually applied for being assigned an UID number, and
- the demographic and biometric of the person in the UID database matches that of the documents submitted at the time of enrolment as well as that of the actual person traced to the address.

Only such an audit can establish the UID database has any more validity than the PDS,

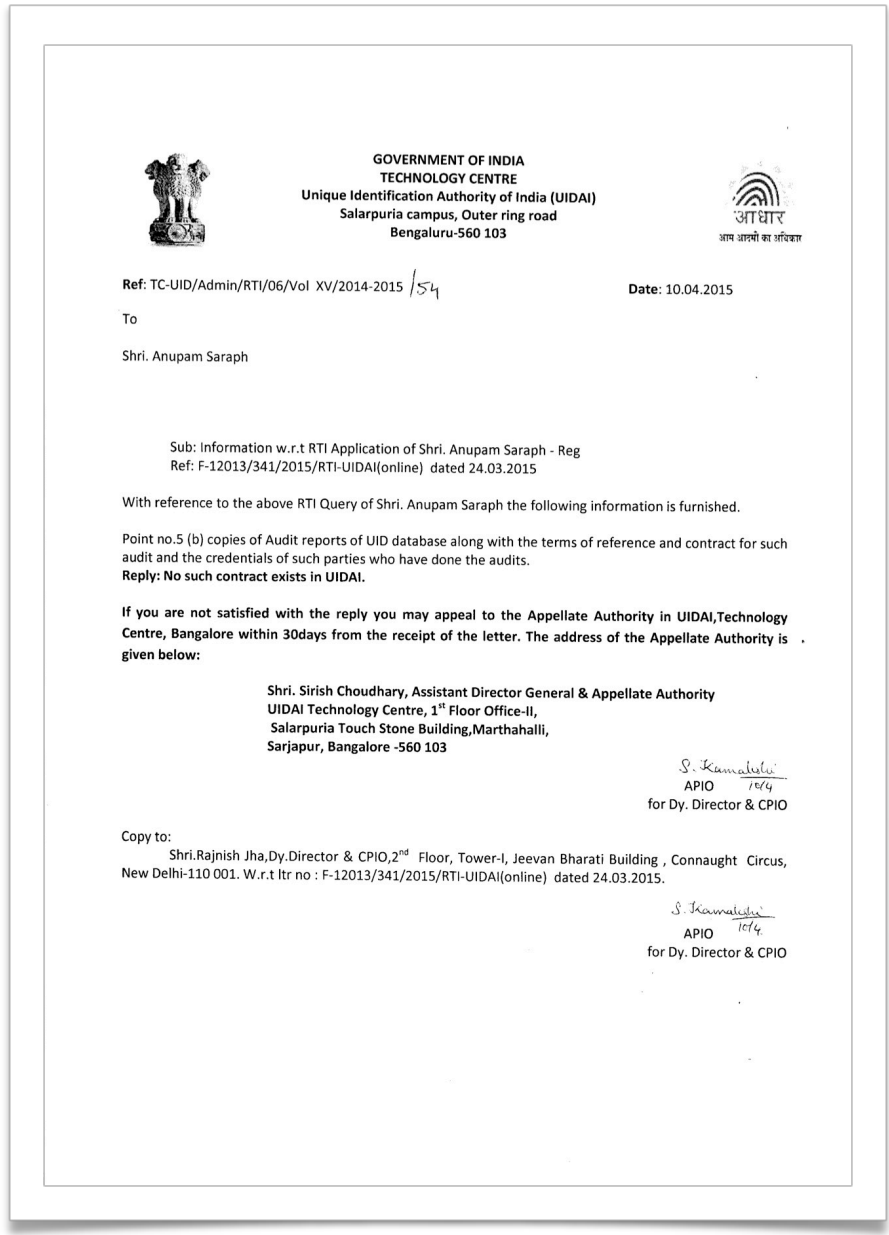

Passport, LPG, EPIC , PAN or other databases already in existence with the government.

An audit of the logs of the use of the UID number can establish faith that in the event of identity thefts, or frauds involving the misuse of an UID number, it would be possible to trace the spread and use of the UID or information associated with the UID and rollback the misuse.

The UIDAI has indicated under a Right to Information request that there has been no contract issued to date to anyone to undertake any audit of the enrolment process or the accuracy of the CIDR (RTI 2015 Figure 5). The UIDAI also indicated that the information about the documents used to issue an UID number are not compiled/not available (Figure 3). This suggests that a future audit of the UID database may be impossible.

De-duplication can be, incorrectly, regarded as an audit. De-duplication, at best, can indicate that there are no duplicate records in the database. Unlike an audit, it cannot vouch that the records correspond to a real person, the same person had actually applied for being assigned a UID number, and the demographic and biometric of the person in the UID database matches that of the documents submitted at the time of enrolment as well as that of the actual person traced to the address.

Sometimes de-duplication using demographic or biometric data is cited as a means to ensure uniqueness (UIDAI 2012c). De-duplication is a process to reject any records with identical biometric and demographic information from the database. Since, the UIDAI outsources the de-duplication to private organizations with proprietary technology, there is no information on the exact process of de-duplication (Diwan 2014). Furthermore, establishing any trust in de-duplication itself necessitates an audit of the de-duplication process itself and also of the stability of the de-duplication process as demographics and biometrics change with time (Ramanathan 2015). Moreover, the UID has also recognized that 1:1 match cannot be obtained by using biometric information and therefore a de-duplication by biometric is impossible (UIDAI 2012e).

The UIDAI has also indicated in response to an RTI question that the CIDR when queried with a biometric data of an individual cannot retrieve a unique record (RTI 2018). It follows that if a unique record cannot be retrieved, there is no possibility of using the biometric to de-duplicate itself or any other database.

The UIDAI provides authentication through licensed private parties, the Authentication Service Agencies (ASA), the KYC Service Agencies (KSA) who service the Authentication User Agencies and the KYC User Agencies (KUA) respectively. This shifts the responsibility of the use of the UID and its associated data away from the UIDAI. Thus, in a possible scenario of misuse, information sought through eKYC for one transaction can be used for multiple transactions without the knowledge of the individual whose information it is. The auditability of the chain of information and its availability to rollback misuse of an ID has never happened.

In the absence of an audit, we can conclude that, the UID database does not establish that every person listed in the database can be found. It does not establish that the demographic and biometric information held in the database must always match any real persons correctly. The UIDAI has no way to distinguish use and misuse of any UID numbers. Not having audited the logs of the use of the UID number, the UIDAI cannot provide any confidence that they can trace and rollback any misuse of any UID number. The UIDAI, therefore, cannot claim any ability to rollback any misuse of information associated with the UID.

*Does the UIDAI allow to see the history of updation of demographic and biometric information associated with the UID number?*

Whenever the demographic or biometric information associated with an UID number is updated, the UIDAI overwrites information associated with the UID number. It does not append the changes (UIDAI 2012b).

There are multiple channels through which the information associated with the UID number can be updated. 'Assisted' modes allow updates in a process similar to the enrolment. Self service modes require the request to be submitted via an Online Portal, Contact Center, Registered Mobile and Physical Post. The requested data may be updated at a later stage at UIDAI's Update back-office by a 'Verifier'. This does not provide for any 2-step verification, or any other parallel verification to ensure the person requesting the update is genuinely the one requesting the update. If the resident data includes mobile number or email, the UIDAI sends an SMS with Update Request Number (URN) to the resident's declared mobile and an electronic notification after every Update. However, any update in mobile and email ID is communicated by way of electronic notification only. No letter is sent to the resident in this case (UIDAI 2012b).

The UIDAI does however, keep a log of the changes for a period of seven years for all fields except date and origin of birth, gender, financial information consent and data sharing consent. For these fields, they retain the history as permanent records (UIDAI, 2012b). The UIDAI policy is not explicit about the accessibility of the log data or the purposes for which it may be used.

For security reasons, the UIDAI requires a one-time-password (OTP) to be sent to the registered mobile number of the individual whose UID information is being updated. In practice, with the mobile numbers being issued using the UID as a proof allows third parties to take control of UID numbers by associating new mobile numbers and updating information linked to them.

It is evident that alteration of records is possible through multiple channels. It is evident from the above, not detailed here for security reasons, that serious changes including alteration of biometric and demographic information in records can happen and ensuring that the resident will not receive updates (UIDAI 2012b). Already 85,67,177 numbers have been deactivated due to a biometric update according to the UIDAI (Yadav 2017).

We can conclude from the above, that while the information associated with the UID number is easily updatable, there is no mechanism in place for the UID number holder to obtain the date of update, previous demographic or biometric entries, or even be guaranteed the genuineness and validity of the current entry.

Without a verification and audit of the processes that generated the UID number, and the database of such UID numbers itself, little trust can be placed in such UID numbers. Without a means to see every change made to the UID numbers they do not inspire confidence in the validity of the information that they can provide.

## Conclusion

The UID enrolment process does not meet the logical requirements for processes to issue identification documents. It does not identify any individual during the act of enrolling. The UIDAI does it certify the identity, age, resident status, address or even the existence of anyone. The number, therefore, cannot serve as a proof of identity, age, resident status, address or even the existence of a real person.

Without any identification of the individual, and any other required attributes like resident status, address or existence, the UID cannot serve as a reliable, impartial and reproducible basis for establishing identity. Without the UIDAI taking responsibility of certifying the identity and other contents of the UID number, the UID cannot serve as a responsible and legally enforceable basis for establishing identity, particularly for use by third parties.

The UIDAI cannot authenticate the UID number and associated information as being the one captured by it during enrolment. It is therefore impossible to distinguish a genuine UID number or card from a fake one.

The UIDAI confuses identification and authentication. Identification is only possible by a person responsible for identification is co-located with the person being identified and responsible for the identification. UIDAI misleads users to believe that authenticating a biometric or demographic field with the UIDAI's CIDR is identification. Not only does this process have no party co-present for identification, it has no party responsible for identification. A query of the UIDAI's CIDR cannot distinguish between stored or fake biometric or demographic data submitted to it and a real person present at the point of authentication. It fails to recognise that authentication is merely the proof of possession of a key, not what identification requires: a proof of being the legal and valid owner of the key.

Furthermore the UIDAI cannot retrieve a unique record when queried by an individual biometric. It, therefore, cannot serve as a unique ID or uniquely identify anyone.

The UIDAI has no means to regulate, control, restrict, track or trace the use of the UID number or the information associated with it. The UIDAI does not provide persons holding a UID number the ability to regulate, control, restrict, track or trace the use or their number or the information associated with it. The UID number, therefore, does not meet the logical requirements to be used as an identification document.

Without authentication of the UID number used for any process, genuineness of the process it is used for cannot be established. Without defining the identification process and the parties responsible for identification, there is no identification. In the absence of a means to regulate and restrict the use of the UID number, misuse or misidentification cannot be prevented.

The UID numbers have not been audited to verify the demographic and biometric information associated with is true, correct or associated with real persons who applied for the number. The UID number does not maintain or provide a history of the updates of information associated with a UID number. The UID number, therefore, does not meet the logical requirements to be used as a valid identification document.

The UID number, therefore, fails to meet the logical requirements for issue, use and validity of identification documents to serve as an identity or address document. It is neither a proof of identity, proof of address or even a proof of existence.

Without a verification and audit of the processes that generated the UID number, and the database of such UID numbers itself, little trust can be placed in such UID numbers. Without a means to see every change made to the UID numbers they do not inspire confidence in the validity of the information that they can provide.

Furthermore, the use of UID numbers fails to ensure justice to those who may lack an identity, be the victims of identity crimes, or even protect persons from violations of their human rights. Unlike other IDs the UID has no ability to trace misuse and deny fraud. No system promising justice can hope to deliver justice to persons who are claimed to be identified using a UID number or all their documents pertaining to their place of residence, their property ownership, their birth and marriage records, their bank accounts, their tax returns or for that matter any transactions depend on the UID. As the UID rewrites voters lists, allows the selective civil death of constituents and also allows the creation of shell bank accounts for corruption and money laundering, it destroys the national interests like no other ID. It exposes India to losing its republic, democratic, or sovereign status with the corruption of its citizen registries and other associated databases vital to independence and self-rule.

The championing and acceptance of the UID number internationally would result in risking identification required for any process within or outside India. It opens up the world to organized colonization, corruption and destruction of databases used for governance, security and financial transactions.

There is no ground for the widespread use of the UID number to replace existing KYC documents and processes and exposing such organizations to colonization, corruption and destruction of their existing databases. There is no ground for exposing individuals to exclusion by service providers authenticating service delivery based on the UID exposed or to deficiency of service provision, service delivery leakages, and servicing fraudulent claims.

*We conclude that the UID number fails to establish identity, undertake identification and build confidence to impartial, reliable and valid identification that can allow the impartial arbitration of the identification of individuals.*